\documentstyle[prl,aps,multicol,epsf]{revtex}
\input epsf

\newcommand{\D}{{\rm d}}
\newcommand{\I}{{\rm i}}
\newcommand{\Tr}{{\rm Tr}}

\begin{document}


\title{Andreev tunnelling in quantum dots: A slave-boson approach}

\author{ P. Schwab$^{a}$, R. Raimondi$^{b}$}
\address{
$^{a}$ Istituto di Fisica della Materia e Dipartimento di Fisica, 
Universit\'a degli Studi di Roma ``La
Sapienza'', Piazzale Aldo Moro 2, I-00185 Roma, Italy}
\address{ $^{b}$ Istituto di Fisica della Materia e  
Dipartimento di Fisica E. Amaldi, Terza Universit\`a
degli Studi di Roma, Via della Vasca Navale 84, I-00146 Roma, Italy}

\date{\today}

\maketitle

\begin{abstract}
We study a strongly interacting
quantum dot connected to a normal and to a superconducting lead.
By means of the slave-boson technique we investigate the low
temperature regime and discuss electrical transport through the dot.
We find that the zero bias anomaly in the current-voltage characteristics
which is
associated to the occurance of the Kondo resonance in the
quantum dot, is enhanced in the presence of superconductivity,
due to resonant Andreev scattering.

\end{abstract}

\pacs{\Pacs{73}{40.Gk}{Tunnelling}
      \Pacs{74}{50.+r}{Proximity effects, weak links, tunnelling phenomena, and Josephson
          effects}}
\begin{multicols}{2}

With the advent of nanotechnology,  recent years have witnessed
 an impressive
experimental activity,  studying various properties of small mesoscopic
structures. In particular, the transport properties of hybrid 
superconducting structures
and the associated Andreev scattering mechanism
have been  investigated intensively after observing several new 
phenomena\cite{Hekking94}. 
Many of these phenomena have been successfully explained  in terms of 
a one-particle
picture essentially based on the BCS theory either via the Bogolubov-de Gennes 
equations or via quasiclassical Green's function methods, as documented by 
various review articles \cite{Hekking94}.

Besides this, electrical transport through small confined regions, where 
electron-electron interactions are strong, 
has also attracted a lot of interest. Such experimental setups,  {\it e.g.}
quantum dots (QDs), allows one 
to investigate in a controlled way the interplay of
 the electron-electron interaction and 
disorder. In particular it has been pointed out that a QD attached to two
metallic leads resembles an impurity level in a metal. As a consequence, even 
when the dot level is far from the Fermi energy of the leads,
transport will occur due to the Kondo effect\cite{Glazman&Ng}. 
This is due to the formation of a
spin singlet between the impurity level and the conduction electrons, which
gives rise to a quasiparticle peak at the Fermi energy in the dot 
spectral function.  This suggestion  has been explored theoretically by
several  groups\cite{meir93,Ng93,Yeyati93,Schoeller94,Hettler95}
and lead to the prediction that one should observe a
zero-bias anomaly in the current voltage characteristics. 
Such an anomaly has been indeed observed in different QD systems
\cite{Ralph94,Goldhaber97}. 

In a recent letter \cite{Fazio98}, it has been suggested that, if the QD 
is coupled to two different types of leads, {\it i.e.}, a normal  
and a superconducting
lead, resonant  Andreev tunnelling yields a stronger zero-bias anomaly
with a broader temperature region where the effect occurs. 
In the analysis of  \cite{Fazio98}
an  approach  based on the equation-of-motion method  which is
mainly valid at high temperature was used.
In this paper, we extend the analysis of \cite{Fazio98} to the
extreme low temperature regime. To this end we use slave-boson mean
field theory.
This approach has been successfully applied to the low temperature
properties of a Kondo impurity in presence of normal\cite{Coleman84,Read83}
as well as for superconducting conduction electrons \cite{Borkowski94}. 
Despite its simplicity, this method captures the main physical aspects
of the Fermi liquid regime at low temperatures, {\it i.e.}, the formation
of a many-body resonance  at the Fermi energy.
For this reason it presents a convenient framework in which to
study the interplay between Andreev scattering and Coulomb interactions.

We model the N-QD-S system with the Hamiltonian 
\begin{equation}
H=H_{\rm N} +H_{\rm S}+H_{\rm D}+H_{\rm {T,N}}+H_{\rm {T,S}}
\label{model}
\end{equation}
where 
$
H_{\rm N}=\sum_{{\bf k} \sigma} \epsilon_{\bf k} c^{\dagger}_{{\rm N},\bf k \sigma}
 c_{{\rm N},{\bf k} \sigma}
$
, 
$
H_{\rm S}=\sum_{{\bf k} \sigma} \epsilon_{\bf k} c^{\dagger}_{{\rm S},{\bf k} \sigma} c_{{\rm S},{\bf k} \sigma}+
\sum_{\bf k} (\Delta c^{\dagger}_{{\rm S},{\bf k} \uparrow} c^{\dagger}_{{\rm S},-{\bf k} \downarrow}
+ c.c.)
$
and
$
H_{\rm D}=\epsilon_{d} d^{\dagger}_{\sigma}d_{\sigma}+Un_{d \uparrow}
n_{d \downarrow}
$
are the Hamiltonians of the normal lead, the superconducting lead 
($ \Delta $ is the superconducting gap) and the dot respectively.
The single particle energy $\epsilon_{\rm d}$ is double degenerate in the 
spin index $\sigma$ and the interaction is included through the on-site 
repulsion $U$.
The position of the dot level can be modulated by an external gate voltage.
Tunnelling between the leads and the dot is described by 
$
H_{\rm {T,\alpha}}=
\sum_{{\bf k} \sigma} (V_{\alpha} c^{\dagger}_{\alpha,{\bf k} \sigma} d_{\sigma}
+ c.c. )
$
where $\alpha ={\rm N},{\rm S}$ and $V_{\alpha }$ is the tunnelling amplitude.
 For simplicity we assumed $V_{\alpha }$ independent from
${\bf k}$ and $\sigma$.  

In the following, we consider 
the on-site repulsion $U$ is infinite, so 
processes where the dot level  is doubly occupied  are excluded.
This condition allows to apply the slave boson technique\cite{Coleman84}.
The dot level is represented as
$d^\dagger_\sigma = f^\dagger_\sigma b$, 
where the fermion $f_\sigma$   and the boson $b$ describe the singly occupied 
and  empty  dot states.
 Since the dot is either empty or singly occupied, the constraint
$
b^\dagger b + \sum_\sigma f^\dagger_\sigma f_\sigma =1 
$
has to be fulfilled.

In the mean field approximation, the operator $b$ is replaced by a $c$-number
$b_0$, and the constraint is fulfilled only on average.
This is achieved by introducing  a chemical potential
$\lambda_0$ for the pseudo particles.
Notice that one ends up
with a non interacting-like problem with renormalized parameters,
{\it i.e.}, an energy shift for the dot level
$\epsilon_{\rm d} \rightarrow \epsilon_{\rm d} +\lambda_0=\tilde \epsilon_{\rm d}$ and a multiplicatively renormalized tunnelling
 $V_\alpha \rightarrow b_0 V_\alpha$.

We discuss the mean field equations and its solution
first in equilibrium and then generalize to non-equilibrium. 
We start from the impurity part of the free energy, which in 
presence of both normal and  superconducting leads is given by
\begin{equation} \label{sb2}
F= - T\sum_{\epsilon_n} \Tr \ln [ 
\I \epsilon_n \hat\sigma^0 - \tilde \epsilon_{\rm d} \hat\sigma^z -b_0^2
  \hat \Gamma(\I \epsilon_n) ]
+\lambda_0 b_0^2 +\epsilon_{\rm d} - \mu 
,\end{equation}
where $\epsilon_n$ is a fermionic Matsubara frequency,
${\hat {\sigma}}^i$ are the Pauli matrices, and 
\begin{eqnarray}\label{sb3}
\hat \Gamma(\I\epsilon_n )&= & \sum_{{\bf k}, \alpha }
 |V_\alpha|^2 \hat \sigma^z \hat g_{{\bf k}, \alpha }(\I \epsilon_n) \hat \sigma^z
\end{eqnarray}
with $\hat g_{{\bf k}, \alpha }$ being the Green's function of the
lead $\alpha$. 

By minimizing the free energy with respect to $\lambda_0$ and $b_0$
we find the equations
\begin{eqnarray} \label{sb6}
b_0^2 + T\sum_{\epsilon_n} \Tr \left[ \hat {\cal G}(\I \epsilon_n)
 \hat \sigma^z  \right] &=& 0, \\ 
\label{sb7}
b_0 \lambda_0 + b_0T \sum_{\epsilon_n} \Tr \left[ \hat{\cal G}(\I \epsilon_n)
 \hat \Gamma(\I \epsilon_n) \right] &= &0  
,\end{eqnarray}
which have to be solved self-consistently.
$\hat{\cal G}(\I \epsilon_n)$ is the pseudo fermion Green function
given by
$\hat{\cal G}(\I \epsilon_n) =
[\I \epsilon_n \hat\sigma^0 - \tilde \epsilon_{\rm d} \hat\sigma^z -b_0^2
  \hat \Gamma(\I \epsilon_n)]^{-1}$
.

Before presenting a numerical solution of the above equations, it is useful to
get some insight from an approximate analytical treatment.
The first equation, eq.(\ref{sb6}), is the constraint. Since the pseudo fermion level is
at maximum singly occupied, the renormalized level is above the Fermi energy.
In the Kondo limit, where the occupancy is nearly one, it is found
$0 < \tilde \epsilon_{\rm d} < b_0^2(\Gamma_{\rm N} + \Gamma_{\rm S}) $, {\it i.e.} $\lambda_0 \approx | \epsilon_{\rm d} |$ and
$\tilde \epsilon_{\rm d} \approx 0$.
The renormalization of the tunnelling amplitude is determined from eq.(\ref{sb7}). A trivial
solution $b_0=0$ always exists. The solutions which minimize the free energy, however, are those 
with $b_0 \ne 0$. 
By introducing a flat density of states in the leads and the tunnelling rates
$\Gamma_\alpha= \pi N_{0\alpha}|V_\alpha|^2$, 
the elements of the matrix
$\hat \Gamma (\I \epsilon_n )$ are
$\Gamma_{11}=\Gamma_{22}=-\I\gamma_1$ and
$\Gamma_{12}=\Gamma_{21}^*=\gamma_2$, where
\begin{equation} 
\label{sb8}
\gamma_1  = {\rm sign}(\epsilon_n) \Gamma_{\rm N} 
+\Gamma_{\rm S} {\epsilon_n \over \sqrt{\epsilon_n^2 + |\Delta|^2} },~ 
\gamma_2  = \Gamma_{\rm S}
 { \Delta \over \sqrt{\epsilon_n^2 + | \Delta|^2} }
.\end{equation}
Restricting ourselves to zero temperature, we replace the Matsubara 
sum in eq.(\ref{sb7}) by an integral and obtain 
\begin{equation}
\label{sb12}
|\epsilon_{\rm d} | = 4 \int_0^W {\D \epsilon \over 2 \pi } 
{ \gamma_1 ( \epsilon + b_0^2 \gamma_1 ) + b_0^2 | \gamma_2|^2  \over
( \epsilon+ b_0^2 \gamma_1 )^2 + b_0^4 | \gamma_2 |^2}
,\end{equation}
where $W$ is a cut-off of order of the band width.
We simplify the integral by approximating $\gamma_1 $ and $\gamma_2$ as
\begin{equation}
\label{sb13}
\gamma_1 =  \left\{ \begin{array}{ll}
\Gamma_{\rm N} & {\rm for }\,\,\, \epsilon < \Delta  \\
\Gamma_{\rm N}+ \Gamma_{\rm S} & {\rm for }\,\,\, \epsilon > \Delta
\end{array} \right. 
,~ \gamma_2 = \left\{
\begin{array}{ll}
\Gamma_{\rm S} & {\rm for} \,\,\, \epsilon < \Delta \\
0        & {\rm for} \,\,\, \epsilon > \Delta 
\end{array} \right.
\end{equation}
The result is
\begin{eqnarray}
\label{ab15}
|\epsilon_{\rm d}|  &= & {\Gamma_{\rm N} \over \pi }
             \ln{ (\Delta + b_0^2 \Gamma_{\rm N})^2 + b_0^4\Gamma_{\rm S}^2 
  \over
                   b_0^4( \Gamma_{\rm N}^2+\Gamma_{\rm S}^2) } 
\nonumber\\
&&
+{\Gamma_{\rm N}+\Gamma_{\rm S} \over \pi }
\ln{ W^2 \over (\Delta +b_0^2 \Gamma_{\rm S} +b_0^2 \Gamma_{\rm N} )^2  }
,\end{eqnarray}
where we neglect a term proportional to $\Gamma_{\rm S}$, but without 
any logarithmic factor.
For small superconducting gap, {\it i.e.} $\Delta $ much smaller than
the Kondo 
temperature which is given by 
$T_{\rm K} = b_0^2  \Gamma_{\rm N}+ b_0^2 \Gamma_{\rm S}$,
$\Delta$ is negligible. One can then easily solve eq.(\ref{ab15}) for $b_0^2$ and obtains the
result for two normal leads with total tunnelling rate $\Gamma_{\rm N}+\Gamma_{\rm S}$: 
\begin{equation}
\label{sb16}
b_0^2(\Gamma_{\rm N}+\Gamma_{\rm S}) = W \exp\left( -{\pi \over 2}{|\epsilon_{\rm d}|\over \Gamma_{\rm N} + \Gamma_{\rm S} }
\right) 
.\end{equation}
In the opposite limit, where $\Delta$ is much larger than the Kondo temperature, we find
\begin{equation}
\label{sb18}
b_0^2 \sqrt{ \Gamma_{\rm N}^2 + \Gamma_{\rm S}^2 } = W \exp \left(  -{\pi \over 2}
{ |\epsilon_{\rm d}| - ( 2\Gamma_{\rm S} /  \pi ) \ln(W / \Delta) \over \Gamma_{\rm N} } \right)
.\end{equation} 
The results agree qualitatively with what we expect from scaling arguments
for the Anderson model\cite{haldane78}.
In the perturbative regime, a logarithmic correction to $\epsilon_{\rm d}$ has been 
found. 
In the case of a large gap scaling due to the superconducting electrons stops
at energies of the order $\Delta$, giving rise to a finite renormalization of 
$\epsilon_{\rm d}$, as seen in eq.(\ref{sb18}). A similar shift in the dot level has also been reported within the equation-of-motion
approach of ref.\cite{Fazio98}.
In case of a small gap, the superconducting lead contributes to scaling down 
to low energies, where one enters the strong coupling regime.
Presumably, the fixed point is still reached for energies of the order of 
$T_{\rm K}$, much greater than $\Delta$, so that the Kondo temperature 
does not depend on $\Delta$, compare eq.(\ref{sb16}).

Notice that in presence of normal electrons, we always find a non-trivial solution of the mean field equations. This is
to be contrasted with the case of superconducting electrons only, $\Gamma_{\rm N}=0$, where for large gap
only the solution $b_0=0$ exists, and there is no Kondo effect\cite{Borkowski94}.

\begin{figure}
\center
\setlength{\unitlength}{0.1bp}
\special{!
/gnudict 40 dict def
gnudict begin
/Color false def
/Solid false def
/gnulinewidth 5.000 def
/vshift -33 def
/dl {10 mul} def
/hpt 31.5 def
/vpt 31.5 def
/M {moveto} bind def
/L {lineto} bind def
/R {rmoveto} bind def
/V {rlineto} bind def
/vpt2 vpt 2 mul def
/hpt2 hpt 2 mul def
/Lshow { currentpoint stroke M
  0 vshift R show } def
/Rshow { currentpoint stroke M
  dup stringwidth pop neg vshift R show } def
/Cshow { currentpoint stroke M
  dup stringwidth pop -2 div vshift R show } def
/DL { Color {setrgbcolor Solid {pop []} if 0 setdash }
 {pop pop pop Solid {pop []} if 0 setdash} ifelse } def
/BL { stroke gnulinewidth 2 mul setlinewidth } def
/AL { stroke gnulinewidth 2 div setlinewidth } def
/PL { stroke gnulinewidth setlinewidth } def
/LTb { BL [] 0 0 0 DL } def
/LTa { AL [1 dl 2 dl] 0 setdash 0 0 0 setrgbcolor } def
/LT0 { PL [] 0 1 0 DL } def
/LT1 { PL [4 dl 2 dl] 0 0 1 DL } def
/LT2 { PL [2 dl 3 dl] 1 0 0 DL } def
/LT3 { PL [1 dl 1.5 dl] 1 0 1 DL } def
/LT4 { PL [5 dl 2 dl 1 dl 2 dl] 0 1 1 DL } def
/LT5 { PL [4 dl 3 dl 1 dl 3 dl] 1 1 0 DL } def
/LT6 { PL [2 dl 2 dl 2 dl 4 dl] 0 0 0 DL } def
/LT7 { PL [2 dl 2 dl 2 dl 2 dl 2 dl 4 dl] 1 0.3 0 DL } def
/LT8 { PL [2 dl 2 dl 2 dl 2 dl 2 dl 2 dl 2 dl 4 dl] 0.5 0.5 0.5 DL } def
/P { stroke [] 0 setdash
  currentlinewidth 2 div sub M
  0 currentlinewidth V stroke } def
/D { stroke [] 0 setdash 2 copy vpt add M
  hpt neg vpt neg V hpt vpt neg V
  hpt vpt V hpt neg vpt V closepath stroke
  P } def
/A { stroke [] 0 setdash vpt sub M 0 vpt2 V
  currentpoint stroke M
  hpt neg vpt neg R hpt2 0 V stroke
  } def
/B { stroke [] 0 setdash 2 copy exch hpt sub exch vpt add M
  0 vpt2 neg V hpt2 0 V 0 vpt2 V
  hpt2 neg 0 V closepath stroke
  P } def
/C { stroke [] 0 setdash exch hpt sub exch vpt add M
  hpt2 vpt2 neg V currentpoint stroke M
  hpt2 neg 0 R hpt2 vpt2 V stroke } def
/T { stroke [] 0 setdash 2 copy vpt 1.12 mul add M
  hpt neg vpt -1.62 mul V
  hpt 2 mul 0 V
  hpt neg vpt 1.62 mul V closepath stroke
  P  } def
/S { 2 copy A C} def
end
}
\begin{picture}(2339,1403)(0,0)
\special{"
gnudict begin
gsave
50 50 translate
0.100 0.100 scale
0 setgray
/Helvetica findfont 100 scalefont setfont
newpath
-500.000000 -500.000000 translate
LTa
LTb
600 251 M
63 0 V
1493 0 R
-63 0 V
600 334 M
31 0 V
1525 0 R
-31 0 V
600 382 M
31 0 V
1525 0 R
-31 0 V
600 417 M
31 0 V
1525 0 R
-31 0 V
600 443 M
31 0 V
1525 0 R
-31 0 V
600 465 M
31 0 V
1525 0 R
-31 0 V
600 484 M
31 0 V
1525 0 R
-31 0 V
600 500 M
31 0 V
1525 0 R
-31 0 V
600 514 M
31 0 V
1525 0 R
-31 0 V
600 526 M
63 0 V
1493 0 R
-63 0 V
600 609 M
31 0 V
1525 0 R
-31 0 V
600 658 M
31 0 V
1525 0 R
-31 0 V
600 692 M
31 0 V
1525 0 R
-31 0 V
600 719 M
31 0 V
1525 0 R
-31 0 V
600 740 M
31 0 V
1525 0 R
-31 0 V
600 759 M
31 0 V
1525 0 R
-31 0 V
600 775 M
31 0 V
1525 0 R
-31 0 V
600 789 M
31 0 V
1525 0 R
-31 0 V
600 802 M
63 0 V
1493 0 R
-63 0 V
600 884 M
31 0 V
1525 0 R
-31 0 V
600 933 M
31 0 V
1525 0 R
-31 0 V
600 967 M
31 0 V
1525 0 R
-31 0 V
600 994 M
31 0 V
1525 0 R
-31 0 V
600 1016 M
31 0 V
1525 0 R
-31 0 V
600 1034 M
31 0 V
1525 0 R
-31 0 V
600 1050 M
31 0 V
1525 0 R
-31 0 V
600 1064 M
31 0 V
1525 0 R
-31 0 V
600 1077 M
63 0 V
1493 0 R
-63 0 V
600 1160 M
31 0 V
1525 0 R
-31 0 V
600 1208 M
31 0 V
1525 0 R
-31 0 V
600 1242 M
31 0 V
1525 0 R
-31 0 V
600 1269 M
31 0 V
1525 0 R
-31 0 V
600 1291 M
31 0 V
1525 0 R
-31 0 V
600 1309 M
31 0 V
1525 0 R
-31 0 V
600 1325 M
31 0 V
1525 0 R
-31 0 V
600 1339 M
31 0 V
1525 0 R
-31 0 V
600 1352 M
63 0 V
1493 0 R
-63 0 V
600 251 M
0 63 V
0 1038 R
0 -63 V
834 251 M
0 31 V
0 1070 R
0 -31 V
971 251 M
0 31 V
0 1070 R
0 -31 V
1068 251 M
0 31 V
0 1070 R
0 -31 V
1144 251 M
0 31 V
0 1070 R
0 -31 V
1205 251 M
0 31 V
0 1070 R
0 -31 V
1257 251 M
0 31 V
0 1070 R
0 -31 V
1303 251 M
0 31 V
0 1070 R
0 -31 V
1342 251 M
0 31 V
0 1070 R
0 -31 V
1378 251 M
0 63 V
0 1038 R
0 -63 V
1612 251 M
0 31 V
0 1070 R
0 -31 V
1749 251 M
0 31 V
0 1070 R
0 -31 V
1846 251 M
0 31 V
0 1070 R
0 -31 V
1922 251 M
0 31 V
0 1070 R
0 -31 V
1983 251 M
0 31 V
0 1070 R
0 -31 V
2035 251 M
0 31 V
0 1070 R
0 -31 V
2081 251 M
0 31 V
0 1070 R
0 -31 V
2120 251 M
0 31 V
0 1070 R
0 -31 V
2156 251 M
0 63 V
0 1038 R
0 -63 V
600 251 M
1556 0 V
0 1101 V
-1556 0 V
600 251 L
LT0
1438 719 M
180 0 V
538 103 R
-18 5 V
-26 7 V
-26 8 V
-26 9 V
-26 10 V
-25 9 V
-26 11 V
-26 10 V
-26 11 V
-26 10 V
-26 11 V
-26 11 V
-26 11 V
-26 10 V
-26 11 V
-26 10 V
-26 10 V
-26 9 V
-26 10 V
-26 8 V
-25 9 V
-26 8 V
-26 8 V
-26 7 V
-26 7 V
-26 7 V
-26 6 V
-26 6 V
-26 5 V
-26 5 V
-26 4 V
-26 5 V
-26 4 V
-26 3 V
-26 4 V
-25 3 V
-26 3 V
-26 2 V
-26 3 V
-26 2 V
-26 2 V
-26 2 V
-26 2 V
-26 1 V
-26 1 V
-26 2 V
-26 1 V
-26 1 V
-26 1 V
-26 1 V
-25 1 V
-26 0 V
-26 1 V
-26 0 V
-26 1 V
-26 0 V
-26 1 V
-26 0 V
-26 0 V
-26 1 V
-8 0 V
LT1
1438 619 M
180 0 V
2156 487 M
-18 11 V
-26 17 V
-26 19 V
-26 20 V
-26 22 V
-25 22 V
-26 24 V
-26 24 V
-26 25 V
-26 24 V
-26 25 V
-26 25 V
-26 24 V
-26 25 V
-26 24 V
-26 23 V
-26 22 V
-26 22 V
-26 21 V
-26 20 V
-25 19 V
-26 18 V
-26 16 V
-26 15 V
-26 14 V
-26 13 V
-26 12 V
-26 10 V
-26 10 V
-26 8 V
-26 8 V
-26 7 V
-26 6 V
-26 6 V
-26 4 V
-25 5 V
-26 4 V
-26 3 V
-26 3 V
-26 3 V
-26 3 V
-26 2 V
-26 2 V
-26 1 V
-26 2 V
-26 1 V
-26 2 V
-26 1 V
-26 1 V
-26 1 V
-25 1 V
-26 0 V
-26 1 V
-26 0 V
-26 1 V
-26 0 V
-26 1 V
-26 0 V
-26 1 V
-26 0 V
-8 0 V
stroke
grestore
end
showpage
}
\put(1378,619){\makebox(0,0)[r]{$\Gamma_{\rm S} = 2 \Gamma_{\rm N}$}}
\put(1378,719){\makebox(0,0)[r]{$\Gamma_{\rm S} =\hphantom{2} \Gamma_{\rm N}$}}
\put(1378,51){\makebox(0,0){$\Delta/ W $}}
\put(100,801){%
\special{ps: gsave currentpoint currentpoint translate
270 rotate neg exch neg exch translate}%
\makebox(0,0)[b]{\shortstack{$b_0^2$}}%
\special{ps: currentpoint grestore moveto}%
}
\put(2156,151){\makebox(0,0){1}}
\put(1378,151){\makebox(0,0){0.1}}
\put(600,151){\makebox(0,0){0.01}}
\put(540,1352){\makebox(0,0)[r]{1}}
\put(540,1077){\makebox(0,0)[r]{0.1}}
\put(540,802){\makebox(0,0)[r]{0.01}}
\put(540,526){\makebox(0,0)[r]{0.001}}
\put(540,251){\makebox(0,0)[r]{0.0001}}
\end{picture}
\narrowtext
\caption{The mean field parameter $b_0^2$, which is a measure for the
Kondo temperature, as a function of the superconducting gap. For both curves
the energy of the dot level is $\epsilon_{\rm d} = -W/3$ and the total tunnelling
rate
$\Gamma_{\rm S} + \Gamma_{\rm N} = 0.15W$. The Kondo temperature estimated from
 these
parameters is $T_{\rm K}  \approx 0.03W$.}
\label{fig1}
\end{figure}
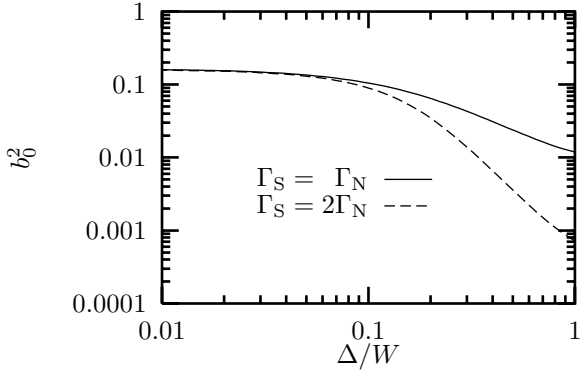
In Fig.\ref{fig1} we report numerical results for $b_0^2$ as a function of $\Delta$.
Instead of a sharp cut-off in the density of states, we used
an exponential, $ \exp(-\epsilon^2/W^2 )$, in our numerics.
The bare $d$-level is
$\epsilon_{\rm d} = -W/3$. For $\Gamma_{\rm N}+\Gamma_{\rm S} = 0.15W$ the Kondo
temperature is approximately
given by $T_{\rm K} \approx 0.03 W $. As long as the gap remains below
this energy scale, $b_0^2$ is almost constant.
It drops quickly for $T_{\rm K} < \Delta < W$,
with a region, where $\log b_0^2 $ decreases linearly in $\log \Delta$, in agreement with eq.(\ref{sb18}).

In a non-equilibrium situation, when a voltage is applied between the two leads,
the mean field parameters cannot be obtained by minimizing the free energy.
However one can derive self-consistency equations following the arguments 
given in ref.\cite{Millis87}. These impose the vanishing of so called
tadpole diagrams order by order in perturbation theory.
The equations read
\begin{eqnarray} \label{sb21}
b_0^2 -\I \int {\D \epsilon \over 2\pi } \Tr \left[ \hat{\cal G }^<(\epsilon )
 \hat \sigma^z \right] &=& 0 \\
\label{sb22}
\lambda_0 b_0 -\I b_0 \int{\D\epsilon \over 2 \pi} \Tr \left[ 
\hat {\cal G}^R(\epsilon) \hat \Gamma^<(\epsilon) + \hat{\cal G}^<(\epsilon) \hat \Gamma^A(\epsilon)
\right] &=&0
,\end{eqnarray}
where the lesser Green's function
$ \hat{\cal G}^<(t, t' ) = \I \langle \phi^\dagger (t')
 \phi(t)    \rangle$
has been introduced,
with 
$\phi=( f_\uparrow, f^\dagger_\downarrow )$.
The lesser and advanced matrix $\hat \Gamma$ is defined in analogy to its equilibrium 
version in eq.(\ref{sb3}).
To obtain $\hat {\cal G}^<$, we use the general relation
$\hat {\cal G}^<= \hat{\cal G}^R \hat \Sigma^< \hat{\cal G}^A$,
where at mean-field level $\hat \Sigma^<= b_0^2 \hat \Gamma^< $ and
\begin{equation}\label{sb23}
{\hat {\Gamma}}^< (\epsilon)=-\sum_{\alpha, {\bf k} } 
|V_\alpha|^2{\hat\sigma}^z
\left[ {\hat g }_{\alpha,{\bf k}}^R(\epsilon) {\hat {f}}_{\alpha} (\epsilon) -
 {\hat {f}}_{\alpha} (\epsilon){\hat {g}}_{\alpha,{\bf k}}^A(\epsilon)\right]
{\hat\sigma}^z.
\label{nonintself}\end{equation}
If the chemical potential of the normal electrode differs by $eV$ from that of the 
superconductor, the matrices $\hat f_\alpha$ have elements
$f_{\alpha,11}=f(\epsilon +eV^{\rm x}_{\alpha})$ and 
$f_{\alpha,22}=1-f(-\epsilon +eV^{\rm x}_{\alpha} )$, with $V^{\rm x}_{\rm N}=V$,
 $V^{\rm x}_{\rm S}=0$ and $f(\epsilon)$ being the
Fermi function.   
Note that the superconducting lead does not contribute
to $\Sigma^< (\epsilon )$ for $\epsilon < \Delta$. 

The solution of the mean-field equations, in  presence  of an external voltage,
can then be obtained along the lines of the equilibrium case.  
As long as $|e V| < T_{\rm K}$ the solution is almost independent of the voltage.
For large voltage, $|eV| \gg T_{\rm K}$, we found the Kondo peak pinned to the
chemical potential at the normal side, {\it i.e.} 
$\tilde\epsilon_{\rm d} \to \tilde\epsilon_{\rm d} -eV$, with a decreased width.

Following ref. \cite{Fazio98},
the Andreev current through an interacting quantum dot is 
\begin{equation}
I= 2\I e \int{\D \epsilon\over 2\pi}\Gamma_{\rm N}
\Tr\left\{ \hat\sigma^z \hat G^R(\epsilon) 
\left[ \hat \Sigma^R(\epsilon) , \hat f_{\rm N}(\epsilon)\right]
\hat G^A(\epsilon)\right\}
,\end{equation}
where $\hat G$ and $\hat \Sigma$ are the Green's function and self-energy of
 the dot electrons.  Within the present
 mean-field approach, $\hat G= b_0^2 \hat {\cal G}$.
Explicitly, the Andreev current is given by
\begin{equation}
\label{current5}
I(V) ={{4e^2}\over h}\int^\infty_{-\infty}\D\epsilon 
{{f(\epsilon -eV)-f(\epsilon +eV)}\over {2e}} G_{{\rm NS}}(\epsilon )
\end{equation}
with 
\begin{equation}
\label{conductance}
G_{{\rm NS}}(\epsilon )=
{4 {(\tilde\Gamma_{\rm N} \tilde\Gamma_{\rm S} )^2 }\over
(\tilde \epsilon^2 -{ \tilde\epsilon}_{\rm d}^2-\tilde \Gamma_{\rm N}^2-\tilde\Gamma_{\rm S}^2)^2
+ 4 \tilde\Gamma_{\rm N}^2 \tilde \epsilon^2 } 
\end{equation}
Here the tunnelling rates 
$\tilde\Gamma_{{\rm S},{\rm N}}=b_0^2\Gamma_{{\rm S},{\rm N}}$,
and
${ \tilde {\epsilon}} = \epsilon (1+b_0^2 \Gamma_{\rm S}/\Delta)$.
Then one recovers the current formula for a non-interacting quantum dot, 
with renormalized parameters which are voltage dependent.
We see that on resonance, when $\tilde\epsilon_{\rm d} \approx 0$ and $\epsilon =0$,
the small renormalization factor $b_0$ drops out and 
the conductance has the same
value as in the non-interacting case\cite{Beenakker92}. 
The peak strength becomes maximal when
 $\tilde{\Gamma}_{\rm N}=\tilde{ \Gamma}_{\rm S}$ with 
$G_{{\rm NS},{\rm max}}=1$,
twice the maximum for a N-QD-N system.

Finally, we want to comment on the reliability of our results. 
The success of slave-boson mean field theory stems from the fact
that it captures the  Fermi-liquid regime at low temperature. 
If the N-QD-S system  scales to a Fermi liquid 
at low temperature, $G_{\rm NS}$ as given in eq.(\ref{conductance}) is
exact in the low
temperature, low voltage limit, with the parameters to be determined.
Within slave-boson mean field theory $\Gamma_{\rm N}$ and $\Gamma_{\rm S}$ 
renormalize equally, although this may no longer be the case when
considering higher order corrections.
For illustration, we estimate the effect of residual quasiparticle interaction
in the limit $\Delta \ll T_{\rm K}$. By assuming an effective quasiparticle
interaction of the form
$H_{\rm int} = \tilde U n_\uparrow n_\downarrow$, we find to first order
in $\tilde U$ no corrections to $\tilde{\Gamma}_{\rm N}$, while, 
as one could have expected, repulsive quasiparticle interaction 
suppresses 
$\tilde{\Gamma}_{\rm S} =b_0^2{\Gamma}_{\rm S}[1-
(\tilde U /\pi T_{\rm K})(\Delta / T_{\rm K})\ln{T_{\rm K}/\Delta}]$. 
 
We studied Andreev tunnelling through a strongly interacting quantum dot, focussing on the
extreme low temperature limit. 
In agreement with a previous study, we
found an enhanced Andreev current at low bias voltage, due to the Kondo effect.
The zero-bias conductance is maximum with the universal value 
$G_{\rm NS} =1 $
when the renormalized tunnelling rates 
$\tilde{\Gamma}_{\rm N}$ and $\tilde{\Gamma}_{\rm S}$ are equal. 
We identified the ratio 
$\Delta/T_{\rm K}$ as an important parameter.
In the case $\Delta \ll T_{\rm K}$,  
the Kondo resonance forms as for two normal leads.
The condition $\tilde \Gamma_{\rm N} = \tilde \Gamma_{\rm S}$ coincides with 
equal bare tunnelling rates ${\Gamma}_{\rm N}={\Gamma}_{\rm S}$. In case of
large gap, quasiparticle interaction suppresses $\tilde{\Gamma}_{\rm S}$,
nevertheless the conductance maximum condition may be achieved with
large bare tunnelling rate ${\Gamma}_{\rm S} > {\Gamma}_{\rm N}$.

We thank R. Fazio for many fruitful discussions.
We acknowledge the financial support of INFM under the PRA-project
"Quantum Transport in Mesoscopic Devices" and the EU TMR programme. 

\end{multicols}
\end{document}